\documentclass[twocolumn,appendixfloats,tighten]{aastex6}

\usepackage{graphicx}	
\usepackage{amsmath}	
\usepackage{amssymb}	
\usepackage{mathtools}
\usepackage{units}
\usepackage{enumitem}
\usepackage{bm}
\usepackage{color}

\def\msun{{\rm\,M_\odot}}

\def\msun{{\rm\,M_\odot}}

\newcommand{\be}{\begin{equation}}
\newcommand{\ee}{\end{equation}}

\def\h2{${\rm\,H_2}$}

\newcommand{\beq}{\begin{equation}}
\newcommand{\beqa}{\begin{eqnarray}}
		 \newcommand{\eeq}{\end{equation}}
\newcommand{\eeqa}{\end{eqnarray}}

\begin{document}

\title{A Common Origin for Low Mass Ratio Events Observed by LIGO and Virgo in the First Half of the Third Observing Run}
\author{Mohammadtaher Safarzadeh\altaffilmark{1,2} }
\altaffiltext{1}{Department of Astronomy \& Astrophysics, University of California, Santa Cruz, CA 95064, USA
\href{mailto:msafarza@ucsc.edu}{msafarza@ucsc.edu}
}
\altaffiltext{2}{Center for Astrophysics | Harvard \& Smithsonian, 60 Garden Street, Cambridge, MA, USA}
\author{Daniel Wysocki\altaffilmark{3}}
\altaffiltext{3}{Leonard E.~Parker Center for Gravitation, Cosmology, and Astrophysics, University of Wisconsin--Milwaukee, Milwaukee, WI 53201, USA}

\begin{abstract}
In its third observing run, the LIGO/Virgo collaboration has announced a potential neutron star-black hole (NSBH) merger candidate, GW190426\_152155. Together with GW190814, these two events belong to a class of binaries with a secondary mass less than $3 \, \msun{}$. While the secondary system in GW190426\_152155 is consistent with being a neutron star with a mass of $1.5^{+0.8}_{-0.5} \, \msun{}$, that of GW190814 is a $2.59^{+0.08}_{-0.09} \, \msun{}$ object and counts as the first confirmed detection of a mass-gap object. Here we argue that these two events could have a common origin as follows: both are formed as NSBH systems; however, the larger escape velocity of a system with more massive primary BH increases the bound fraction of the ejecta material from the supernova explosion leading to the formation of a NS. This bound material forms a disk, which is preferentially accreted onto the NS. This scenario predicts the secondary component mass should correlate with the primary component mass, which is consistent with GW190426\_152155 and GW190814. If this hypothesis is corroborated by upcoming observations, GW190814-like events can be excluded from the binary black hole population when inferring their global characteristics.
\end{abstract}


\section{Introduction}

The LIGO/Virgo collaboration recently released the detected gravitational-wave events in the first half of their third observing run \citep{TheLIGOScientific:2014jea,TheVirgo:2014hva,abbott2020gwtc2,O3apop}
One notable system is the neutron star-black hole (NSBH) merger candidate, GW190426\_152155, with primary (secondary) component masses of $m_1=5.7^{+0.4}_{-2.3}$ ($m_2=1.5^{+0.8}_{-0.5}$) $\msun{}$. 
Although the false alarm rate of this system is the highest ($\mathrm{FAR} \approx 1.4 \, \mathrm{yr}^{-1}$), it is still possible for this event to have an astrophysical origin. 

Together with GW190814 \citep{GW190814} which has primary (secondary) component masses of $m_1=23.2^{+1.1}_{-1.0}$ ($m_2=2.59^{+0.08}_{-0.09}$) $\msun{}$, these two GW events constitute systems with the lowest mass ratios, 
with $q={0.112}_{-0.009}^{+0.008}~(0.26^{+0.41}_{-0.15})$ for GW190814 (GW190426\_152155), where mass ratio is defined as $q=m_2/m_1$. In fact, the very low mass ratio of GW190814 makes this GW event an outlier in the entire population studied so far \citep{O3apop}, with its secondary being the first confirmed detection of a mass-gap object. Mass-gap objects are compact objects with masses between 2--3 $\msun{}$ that are either the most massive NSs or the least massive BHs \citep{Bailyn1998ApJ,Ozel2010ApJ,Farr2011ApJ}.
Is it possible these systems have a common origin?

Different formation mechanisms have been proposed for the formation of GW190814 such as in AGN disks \citep{Yang2020ApJ}, from hierarchical mergers \citep{Safarzadeh2020ApJa,Liu2020arXiv,Wenbin2021}, common envelope evolution \citep{2020ApJ...899L...1Z}, population III stars \citep{Kinugawa2020arXiv} or consistent with being a NS-BH system \citep{Ming-Zhe2020}. \citet{SafarzadehLoeb2020} argued that GW190814 could have formed as an NSBH system; however, due to the presence of a very massive primary BH companion, when the secondary star exploded and left an NS behind, a fraction of the ejecta from this explosion could have remained bound to the binary. This comes as a result of the increased escape velocity of the binary due to the gravitational pull of the massive primary BH companion given by:
\be
v_{\rm esc}=\left(\frac{2GM_{BH}}{a}\right)^{1/2},
\ee
where $a$ is the orbital separation between the newly formed NS and the BH. The bound material forms a circumbinary disk, but it will be the NS that accretes from the disk and grows its mass from about 1.4 to about 2.6 $\msun{}$. In this scenario, the bound fraction of the ejecta material depends on the primary BH mass. Therefore, less massive primary companions would not lead to significant changes to the final mass of the newly born NS. 

In this \emph{letter}, we show that if GW190426\_152155 is indeed a real NSBH GW event, the component masses of these two systems are in line with the above interpretation. The structure of this \emph{letter} is as follows: in \S2, we briefly discuss how to compute the bound ejecta fraction from a supernova explosion depending on the primary component mass of a binary system. In \S3, we show the expected trends from this model and discuss how these two low mass ratio events agree with the model, and in \S4, we discuss the caveats of the models and plans for future works.

\begin{figure*}
\hspace{-0.2in}
\centering
\includegraphics[width=\columnwidth]{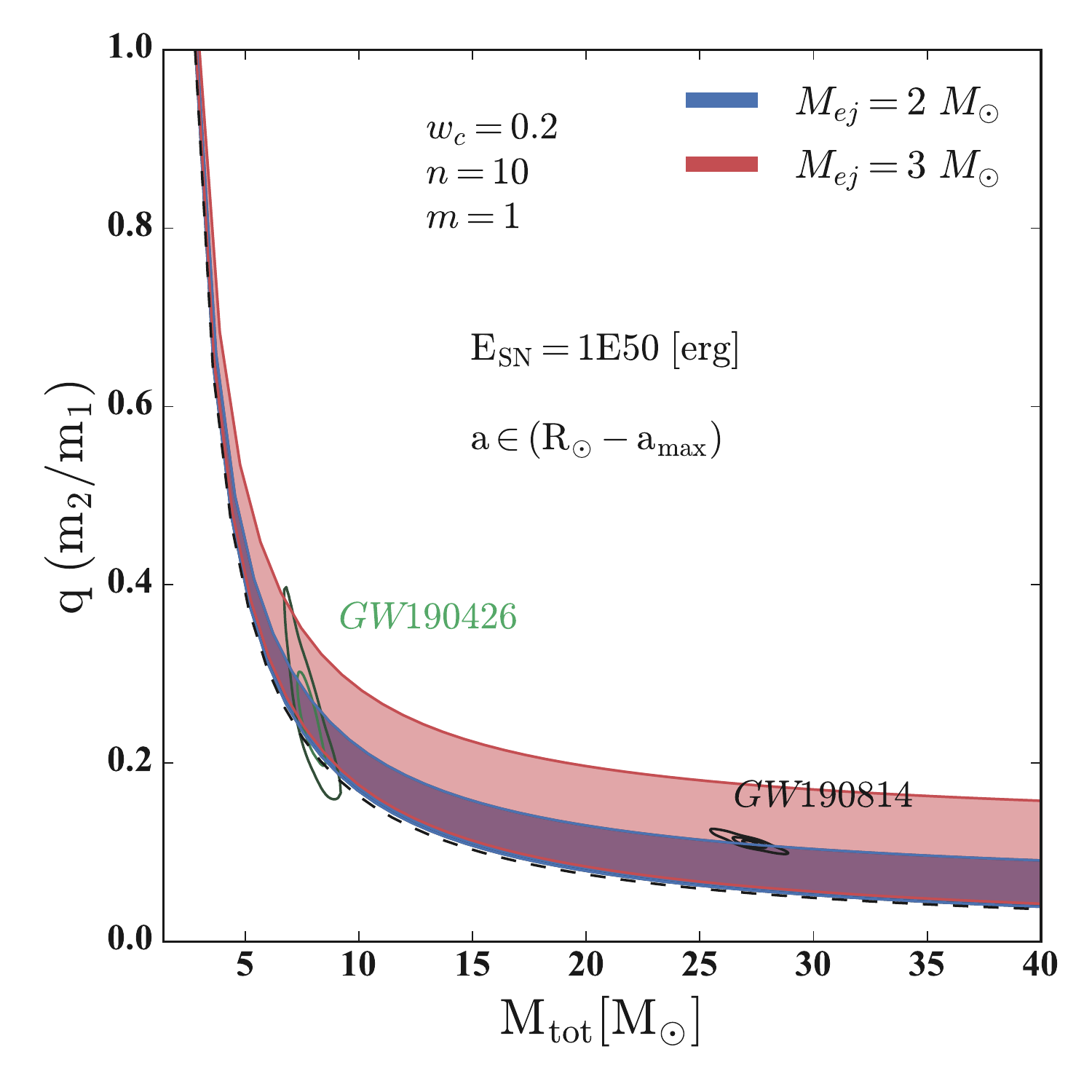}
\includegraphics[width=\columnwidth]{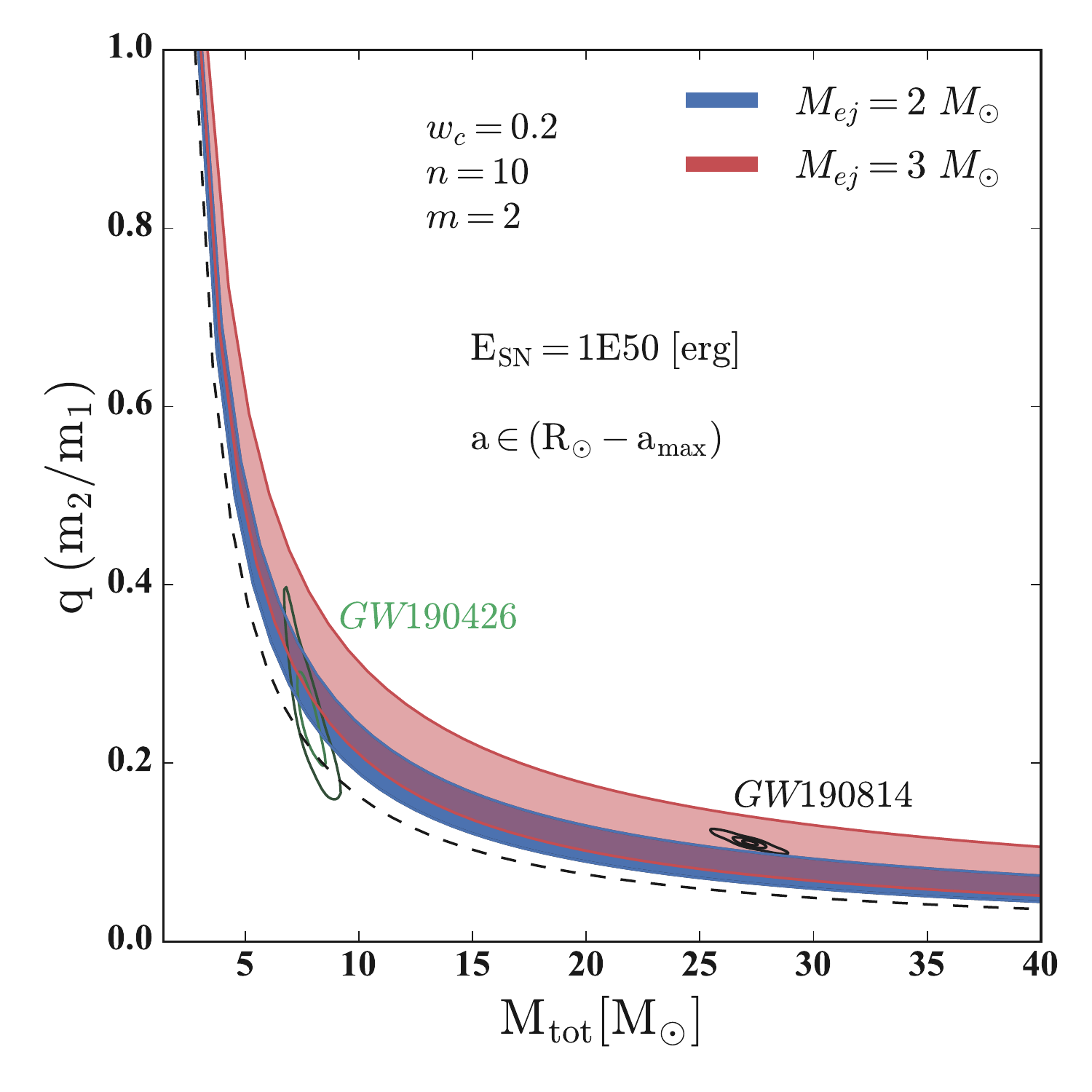}
\includegraphics[width=\columnwidth]{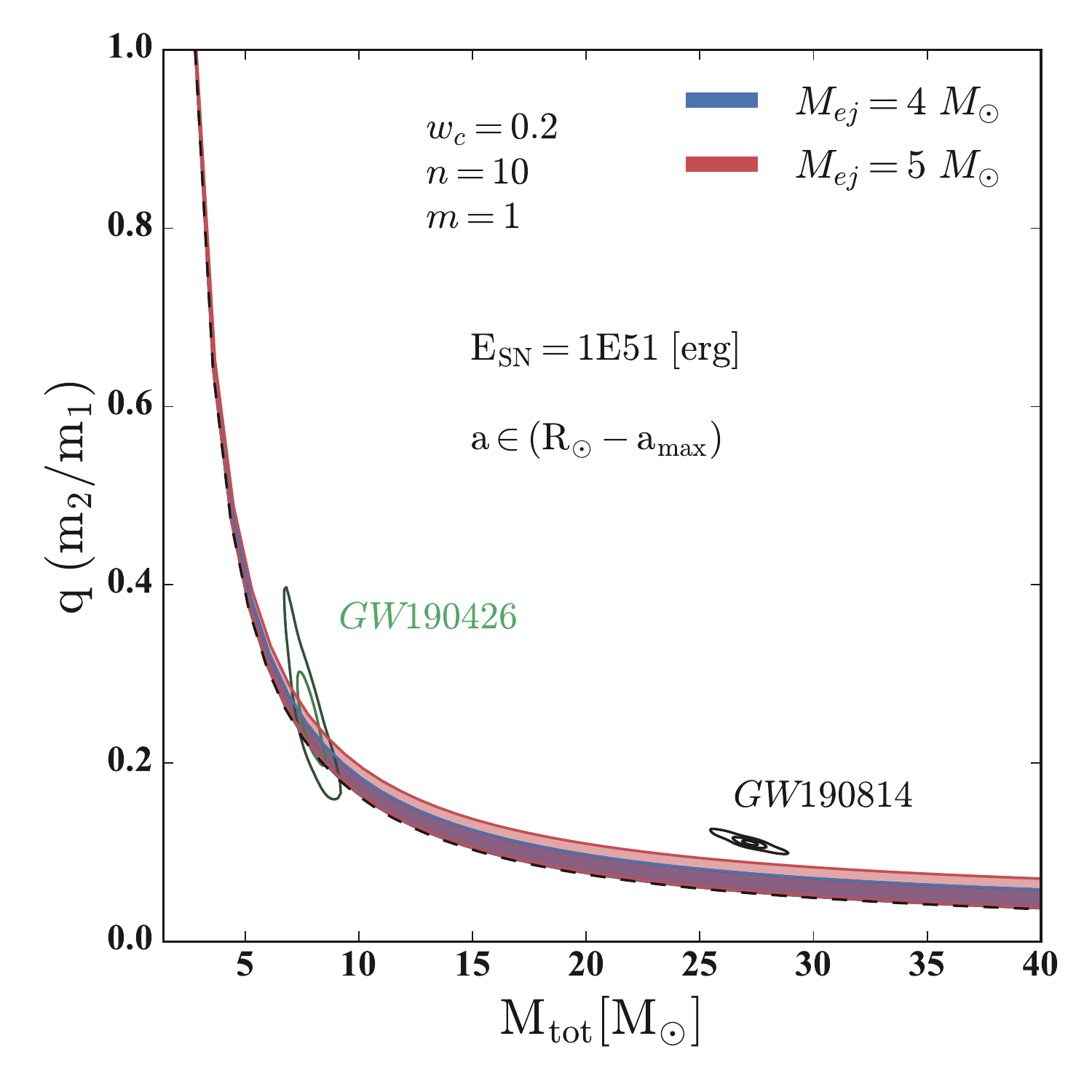}
\includegraphics[width=\columnwidth]{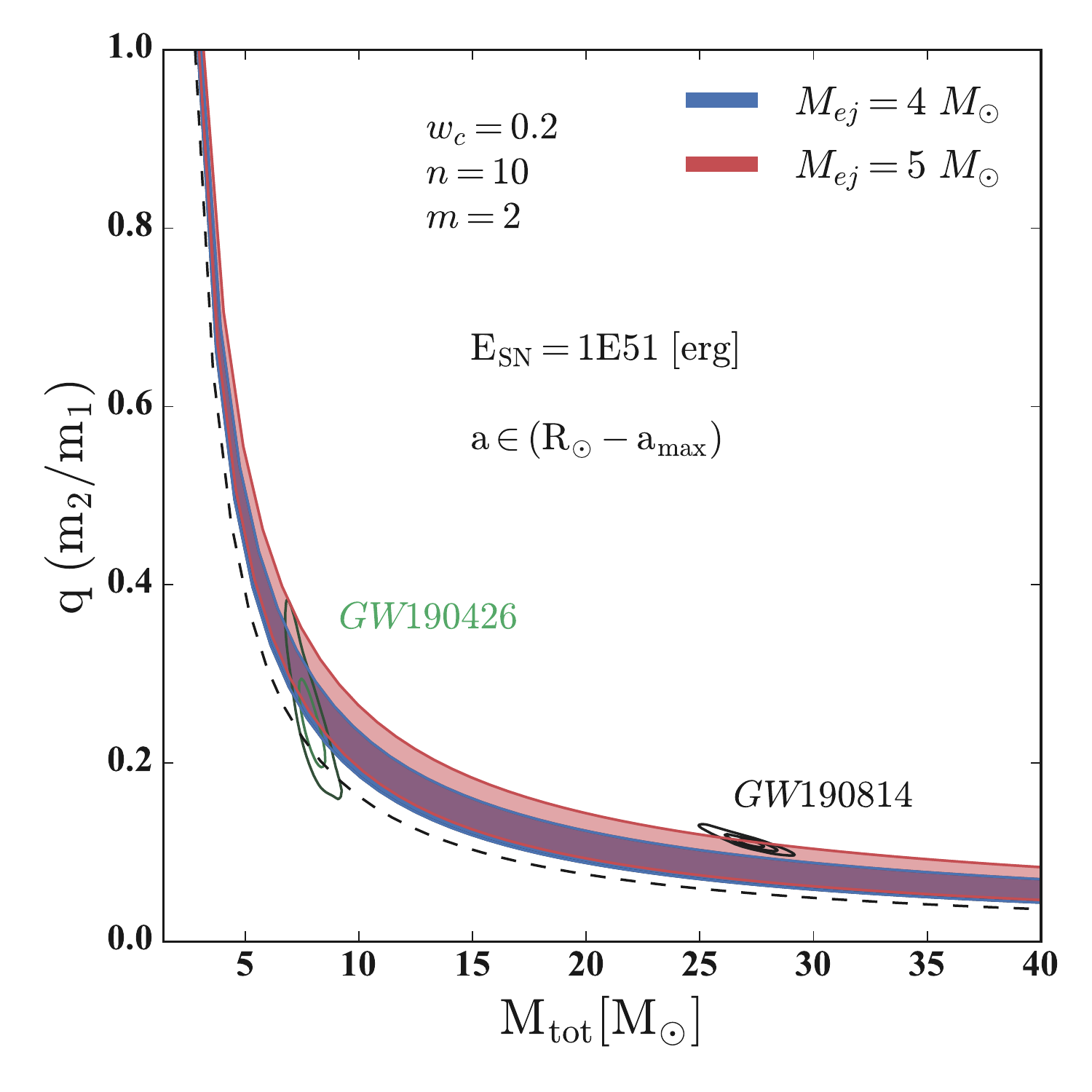}
\caption{The expected correlation between the mass ratio and total mass of NSBH systems that grow in mass through circumbinary accretion disks. The panels show the case assuming a supernova explosion energy of $E_{\rm SN}=10^{50}\rm erg/s$ and ejecta mass of 2 (3) $\msun{}$ shown with blue (red) shaded regions. Top left and top right panels differ on the adopted value for $m$ indicative of the inner slope of the ejecta profile. The envelope in each case comes as a result of varying the orbital separation between the newly formed NS and the primary BH in the binary which we range from minimum of $a_{\rm min}=R_{\odot}$ to $a_{\rm max}$ given by requiring the merging timescale to be less than the age of the universe given the primary BH mass. Bottom panels show the same but assuming $E_{\rm SN}=10^{51}\rm erg/s$ and larger ejecta mass of 4 (5) $\msun{}$ shown with blue (red) shaded regions. In both panels, the posterior estimates of the mass ratio and total masses of GW190426\_152155 and GW190814 based on IMRPhenomNSBH waveform templates are shown with contour plots which are drawn from the GWTC-2. The black dashed lines show the case when there is circumbinary accretion disk, meaning setting $M_{\rm ej}=0$.}
\label{fig_1}
\end{figure*}

\begin{figure*}
\hspace{-0.2in}
\centering
\includegraphics[width=\columnwidth]{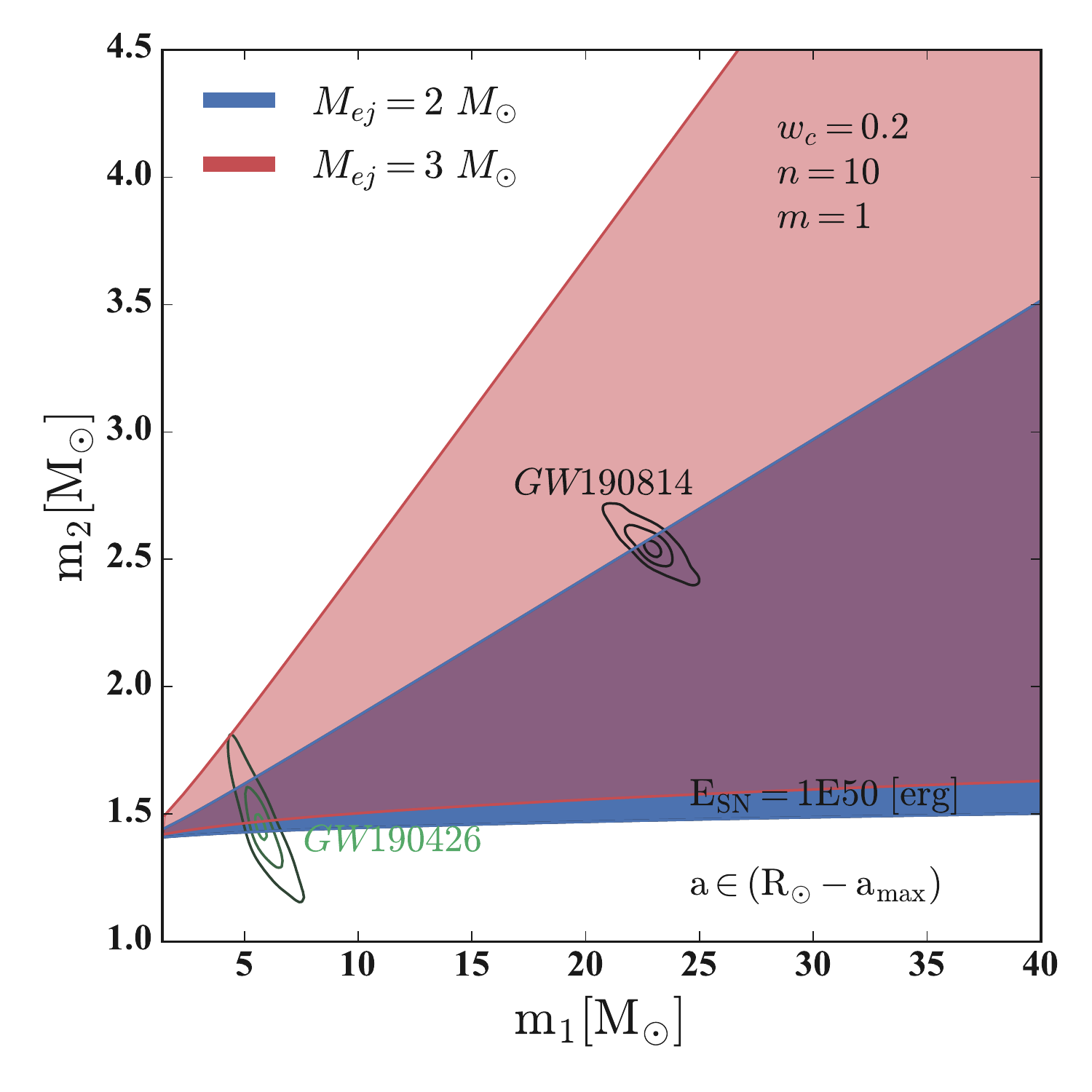}
\includegraphics[width=\columnwidth]{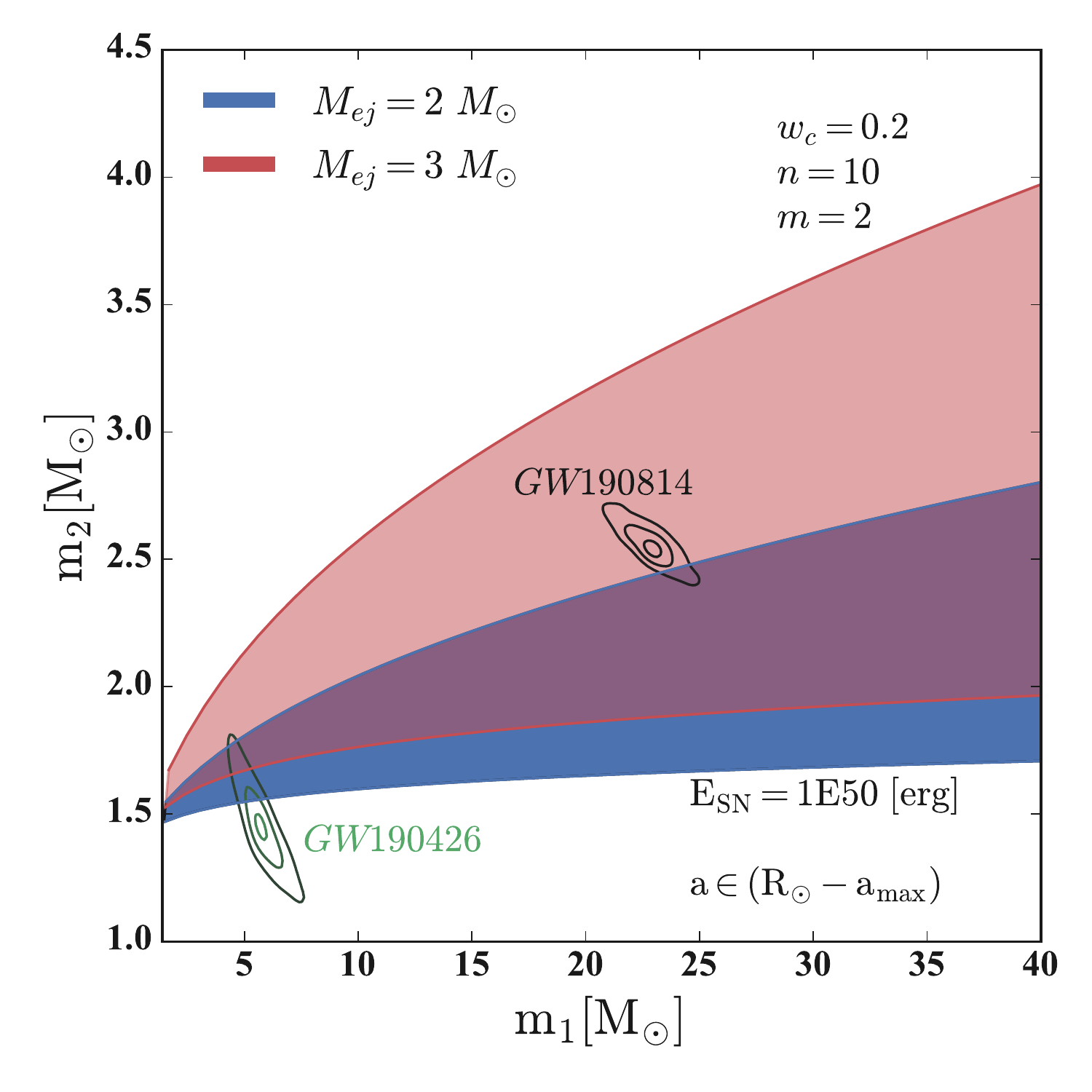}
\includegraphics[width=\columnwidth]{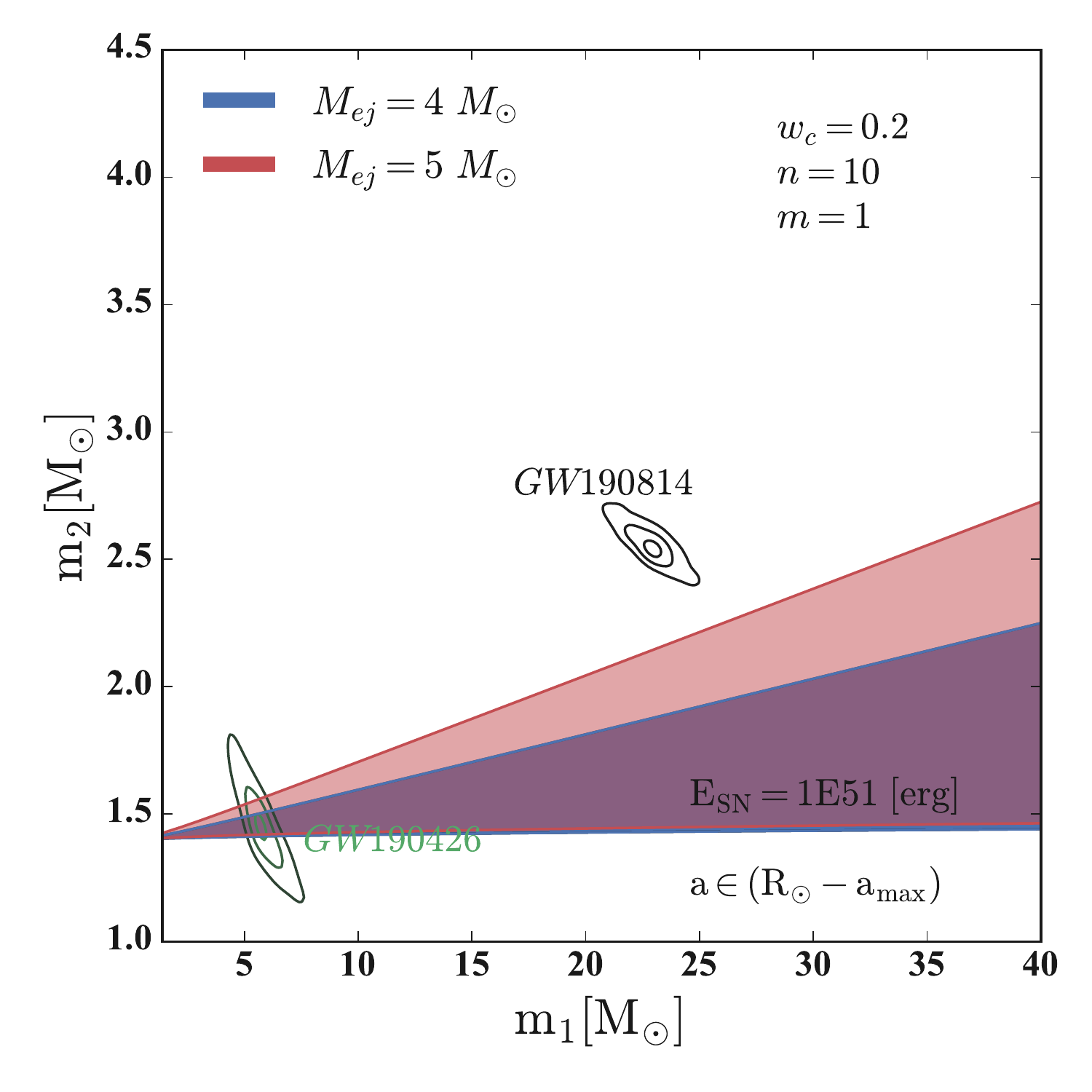}
\includegraphics[width=\columnwidth]{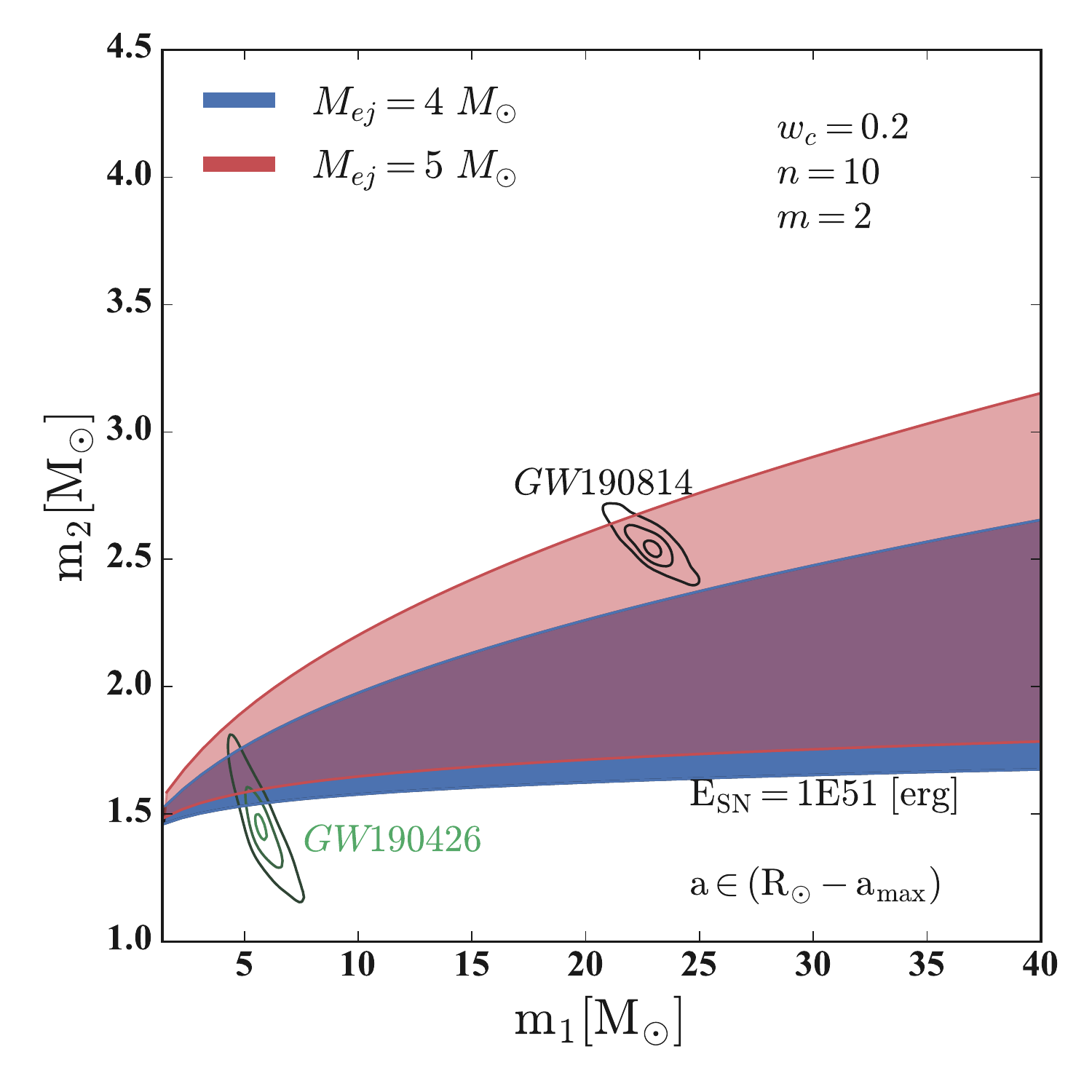}
\caption{The same as Figure \ref{fig_1} but showing the expected correlation between the component masses of the NSBH systems that are processed through circumbinary accretion disks. }
\label{fig_2}
\end{figure*}

\begin{figure}
\centering
\includegraphics[width=\columnwidth]{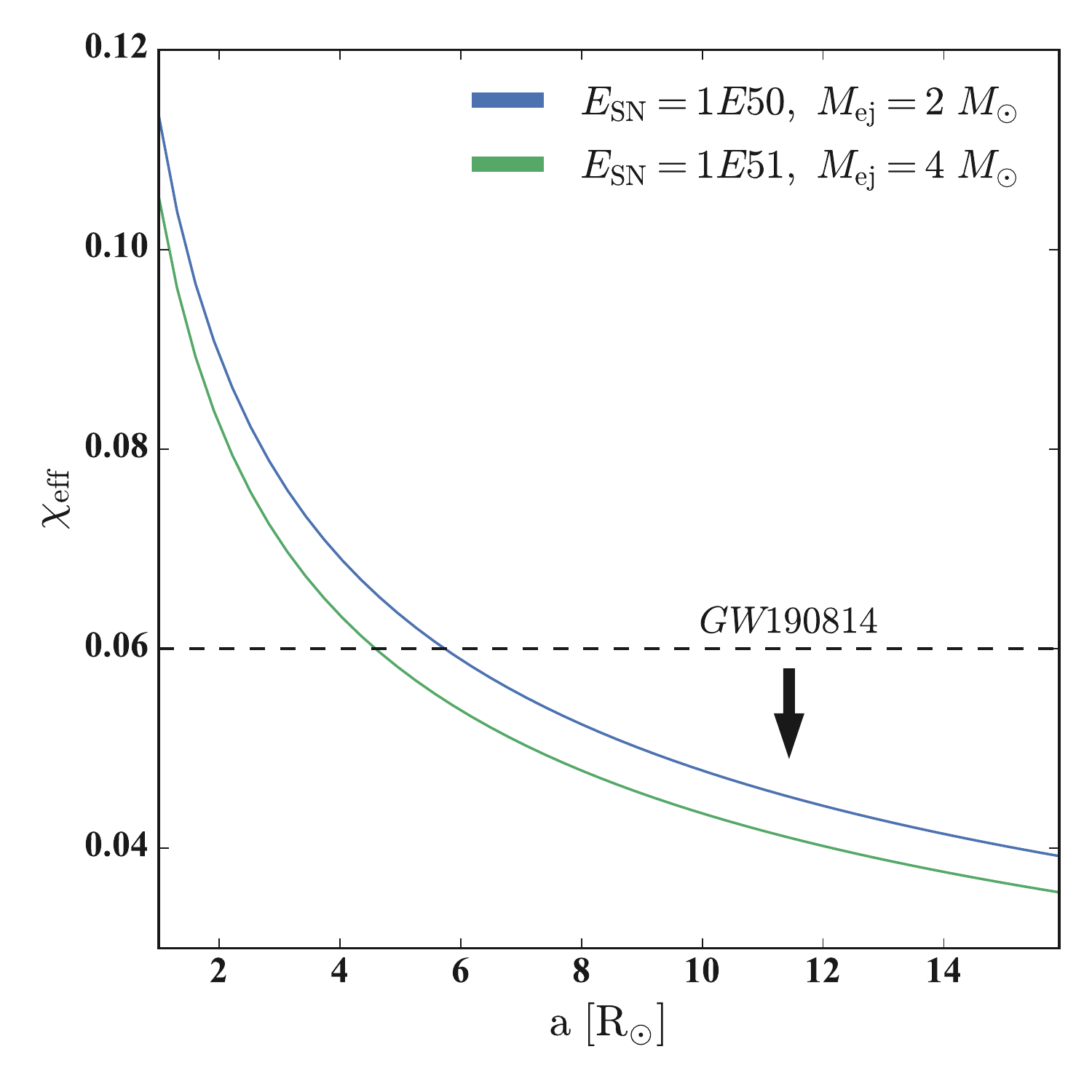}
\caption{The predicted effective spin of GW190814 formed through our proposed channel assuming $m=2, n=10, w_c=0.2$ (ejecta profile parameter). Each line shows the result assuming different SN explosion energy and ejecta mass as a function the assumed separation of the binary with initial masses of 23 and 1.4 $M_{\odot}$ for its components. The dashed line shows the 90\% credible interval on the value of the effective spin for this system which is about 0.06.}
\label{x_eff}
\end{figure}

\begin{figure}
\centering
\includegraphics[width=\columnwidth]{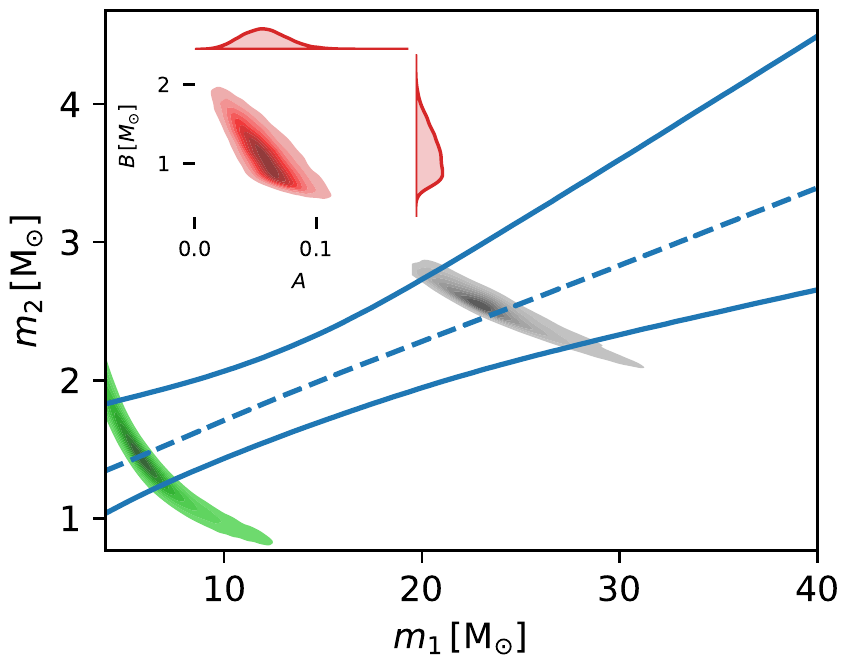}
\caption{Trend lines consistent with the two detections. At each $m_1$, the solid blue lines denote the 5th and 95th posterior percentile, and the dashed blue line shows the median. Inset: posterior distribution on the slope ($A$) and intercept ($B$) for the trend line. Contours show the 2-D joint distribution, while the curves show the respective 1-D marginal distributions.}
\label{trend-line}
\end{figure}

\section{Method}
To estimate the fraction of bound material to a binary after a supernova explosion, we need to know the velocity profile of the ejecta material and estimate what fraction of the total ejecta material is moving with speeds less than the escape velocity of the binary. The details of this calculation are presented in \citet{SafarzadehLoeb2020}, and here we just briefly summarize it.
The formalism is based on the results of \citep{Truelove1999ApJS} and reformulated in \citep{Suzuki2017MNRAS}. The ejecta material is considered to be expanding in a homologous fashion. It has a power-law distribution of density ($\rho$) in velocity space transition with slopes changing from inner to outer part denoted by $m$ and $n$. The ejecta mass with velocity less than $v$ is given by:
\begin{eqnarray}
M(<v)&=&\int_0^{vt}4\pi R^2\rho(t,R)dR
\nonumber\\
&=&
\left\{\begin{array}{llc}
\frac{f_\mathrm{3}M_\mathrm{ej}}{3-m}\left(\frac{v}{w_\mathrm{c}v_\mathrm{ej}}\right)^{3-m}
&\mathrm{, for}&v\leq w_\mathrm{c}v_\mathrm{ej},\\
\frac{f_\mathrm{3}M_\mathrm{ej}}{3-n}
\left[\left(\frac{v}{w_\mathrm{c}v_\mathrm{ej}}\right)^{3-n}
-\frac{n-m}{3-m}
\right]
&\mathrm{, for}&w_\mathrm{c}v_\mathrm{ej}<v.\\
\end{array}\right.
\end{eqnarray}
with a numerical factor $f_{l}$ given by,
\begin{equation}
f_l=\frac{(n-l)(l-m)}{n-m-(l-m)w_\mathrm{c}^{n-l}},
\end{equation}
and the ejecta velocity $v_\mathrm{ej}$ given by:
\be
w_\mathrm{c} v_\mathrm{ej}=\left(\frac{2 f_5 E_{\rm SN}}{f_3 M_{\rm ej}}\right)^{1/2}.
\ee
The parameter $w_\mathrm{c}$ indicates the location of the interface between the inner and outer ejecta in the velocity coordinate.
Following the two zone modeling of the SN ejecta in \citet{Suzuki2017MNRAS} which explored the corresponding parameter ranges of $m \in (0,2)$, $n\in (9,12)$, and $w_c \in (0.1,0.3)$, in this work we set $w_c=0.2$, and $n=10$ while investigating variations in $m$.

The ejected material will subsequently form a circumbinary accretion disk; however, this material will get preferentially accreted onto the smaller compact object of the binary as shown in hydrodynamical simulations of circumbinary accretion disks around asymmetric binaries \citep[e.g., ][]{DOrazio2016MNRAS,Duffell2020ApJ}. The relative ratio of the accreted material between the two components given by the following fitting formula:
\be
\frac{\delta m_2}{\delta m_1}=\frac{1}{0.1+0.9q},
\ee
where $q=m_2/m_1$ is the mass ratio of the binary. Therefore, the final component masses of the binary are increased from their initial value by:
\be
m_2^f=m_2^i+M_{\rm bound}\left(\frac{1}{1.1+0.9q}\right),
\ee
and
\be
m_1^f=m_1^i+M_{\rm bound}\left(1-\frac{1}{1.1+0.9q}\right),
\ee
where $M_{\rm bound}=M(<v_{\rm esc})$. In all these calculations we assume $m_2^i=1.4 \, \msun{}$, and we assume a range for $m_1^i \in (1.4, 40) \, \msun{}$. Moreover, we have applied an efficiency of 80\% to the calculations to account for the velocity boost of the ejecta due to the orbital velocity of the pre-supernova star leading to the formation of the NS \citep{SafarzadehLoeb2020}.

\section{Results}
Figure \ref{fig_1} shows the expected relation between the total mass and mass ratio of NSBH binaries assuming different supernova explosion energy and ejected material mass. 
For each choice, we vary the initial separation of the binary between $a_{\rm min}=1~R_{\odot}$ to $a_{\rm max}$ which is set by requiring the merging timescale due to the emission of GWs to be less than Hubble time ($t_H$) for a binary consist of a $M_{\rm NS}=1.4 \, \msun{}$ NS and a primary BH given by \citep{Peters1964PhRv}:

\be
t_H=\frac{5}{256} \frac{c^5}{G^3}\frac{a_{\rm max}^4}{(M_{\rm BH} M_{\rm NS})(M_{\rm BH}+M_{\rm NS})},
\ee
where $c$ is the speed of light, $G$ is the Newton constant.

Top panels of Figure \ref{fig_1} show the case assuming $E_{\rm SN}=10^{50}\rm erg/s$ and showing two cases of $M_{\rm ej}=2 \, \msun{}$ (blue shaded region) and $M_{\rm ej}=3 \, \msun{}$ (red shaded region).
Bottom panels of Figure \ref{fig_1} show the same but with higher SN explosion energy of $E_{\rm SN}=10^{51}\rm erg/s$ and larger ejecta mass of $M_{\rm ej}=4 \, \msun{}$ (blue shaded region) and $M_{\rm ej}=5 \, \msun{}$ (red shaded region). Top left and top right panels differ on the adopted value for $m$ indicative of the inner slope of the ejecta profile.
The black dashed lines show the case assuming $M_{\rm ej}=0$, meaning turning off this mechanism.  
In each panel, the contours show the posteriors on the mass ratio and total mass of GW190426\_152155 and GW190814 assuming IMRPhenomNSBH waveform model, taken from the GWTC-2 data release \citep{P2000223}. The same is observed if we plot the posteriors from another NSBH waveform, such as SEOBNRv4-ROM-NRTidalv2-NSBH. We note that a higher supernova explosion should be compensated by larger ejecta material mass since a smaller fraction of the ejecta material remains bound to the binary as $E_{\rm SN}$ is increased. The range of supernova explosion energies and the ejecta mass from the pre-supernova He star that we have explored in this work agrees with the expected values from numerical simulations \citep[e.g., ][]{Ertl2020ApJ} which find explosion energies between $(0.2-2)~\times 10^{51}$ erg (see their Figure 14) , and ejecta masses between $(1-5)~M_{\odot}$ (see their Figure 15).

The accreted material will not only increase the mass of the components, but will also increase their spin magnitude. We show the effective spin of GW190814 (defined as the mass weighted projected spin of the components of the binary onto the angular momentum vector of the binary) using the derived fits to the final spin of a compact object knowing its initial and final mass \citep{Bardeen1970,Thorne1974} as a function of the binary's initial separation in Figure \ref{x_eff}. The dashed line is the 90\% credible interval for the effective spin of GW190814, and we present the predictions for two models with different assumptions regarding the SN explosion energy and the ejecta mass. For a large range of the initial binary separation the predicted effective spin is consistent with the observed limit for GW190814. We note that this is largely due to the large mass of the primary component of the system and the fact that most of the ejecta is modeled to be accreted onto the smaller component of the system.

We see that the posterior distribution of these two GW events in mass ratio and total mass agree with the correlation expected to arise if NSBH systems are reprocessed through circumbinary accretion disks. As an exercise, we fit a straight line between these two objects---accounting for selection effects and measurement uncertainty---using the \textsc{PopModels} population inference package \citep{2019PhRvD.100d3012W}. We assume the primary mass distribution obeys a simple Salpeter power-law $m_1^{-2.35}$ \citep{1955ApJ...121..161S}, truncated to $m_{1} \in [4, 50] \msun{}$. $m_{2}$ is assumed to obey $m_{2} = A \, m_{1} + B$, with ($A$, $B$) to be inferred from the data. We only allow models which enforce $m_2 \geq 1 \, \msun{}$ (i.e., $(1 \, \msun{} - B) / A \geq 4 \, \msun{}$) and $m_2 \leq m_1$ (i.e., $B / (1 - A) \leq 4 \, \msun{}$). A description of our method for measuring the posterior is given in Appendix \ref{app:pop-inf}. The posterior rules out any non-positive correlations between $m_1$ and $m_2$, as we show in Figure \ref{trend-line}. However, these strong constraints are a result of our overly simplistic model. Robust measurement of correlations between the two masses would require a much more flexible population model. The assumption that $m_2$ is a pure function of $m_1$ is the main driver of this result, as a non-positively sloped line cannot pass between the two events, but a downward sloping distribution could. Relaxing that assumption, the fixed lower limits on $m_1$ and $m_2$ would also provide a bias towards positive slopes. With only two detections, however, the necessary modeling {d.o.f.} for a robust measurement will leave those {d.o.f.} completely unconstrained, and we will need a much larger sample size to go beyond this proof-of-concept. Additional detections may also ambiguously belong to this formation channel or another due to potentially large uncertainties on the masses. As a result, mixture modeling of the various formation channels will be necessary to infer this channel's properties reliably.

Figure \ref{fig_2} shows the posterior distribution of these two GW events in component masses with the expected trend from the circumbinary accretion disk model. While this is a remapping of the parameter space shown in Figure \ref{fig_1}, we see that 
the mass ratio--total mass plane has a stronger predictive power as the envelope is more constrained between different assumptions going into the model in terms of the supernova energy or ejecta mass.

\section{Discussion}
While the observed location of the two low-mass ratio GW events either in mass ratio and total-mass or primary and secondary component mass plane agree with the circumbinary accretion scenario discussed in this work, more data is required to confidently study the suggested correlations. Moreover, we did not discuss the timescale for the accretion of the material from the circumbinary accretion disk that, in some cases, would require hyper Eddington accretion rates. If we want to avoid such cases, the model would prefer $E_{SN}=10^{50}\rm erg/s$ \citep{SafarzadehLoeb2020}.

One other consequence of this model is that the secondary component's spin magnitude should increase if its mass has increased in the accretion process. For example, \citet{Most2020MNRAS} interpret the secondary of GW190814 to be a highly spinning NS, which would agree with the formation scenario discussed in this work. As such, the spin of the NS with more massive BH companions should be significant, which agrees with the estimates of $0.49^{+0.08}_{-0.05}<a<0.68^{+0.11}_{-0.05}$. We note that in the calculations we preformed to predict the $\chi_{\rm eff}$ of GW190814 through accretion, the spin of the secondary for binaries with effective spin below 0.06 is exactly between $0.45<a<0.62$ or $0.5<a<0.66$ (for the two models shown in Figure \ref{x_eff}) which agrees with the values reported in \citet{Most2020MNRAS}.

We emphasize that this correlation is unique to this proposal. Formation of massive NSs or mass-gap object through fall-back accretion is also predicted to take place \citep[e.g., ][]{Sukhbold2018ApJ}. Although in the fall-back accretion scenario the more massive secondary is expected to have a larger spin magnitude similar to the model discussed in this work, in the fall-back model the mass of the secondary should be independent of its companion primary BH mass. Therefore, with future data, the evidence for or lack thereof a correlation between the primary and secondary mass can differentiate between the circumbinary accretion model that we advocate in this work and the fall-back model. Alternatively, detection of a merger between two mass-gap compact objects would not be easily accommodated within the model discussed in this work. Likewise would be a merger between a mass-gap object with a NS. Such GW events would point to the formation of mass-gap objects directly from the fall-back supernova and not accretion from a circumbinary disk. Moreover, in this work we have not modeled the impact of the NS progenitor’s gravity itself on the ejecta material, but regardless the material would get preferentially accreted on to the NS in a highly asymmetric binary.

In this work, we did not extend this proposal to explain GW190425 \citep{GW1904252020ApJ} since the event is likely to be a binary neutron star merger despite its disputed origin \citep[e.g., ][]{Safarzadeh2020ApJ}.
However, if the system is an NSBH, the implied low mass of the BH in this system would not be large enough to affect the ejecta from the formation of the NS.
Therefore the component masses of the system should have remained close to their birth value.

We also did not extend this proposal to explain GW190412 \citep{2020arXiv200408342T}, which is an asymmetric binary black hole (BBH). BHs are capable of being born with a much wider range of masses than NSs. So while they could, in principle, undergo this same process, the effect will be washed out by the uncertain natal mass.

We also note that if this scenario is at work and is confirmed with future data, GW events similar to GW190814 can be safely excluded from the BBH population when inferring the global characteristics of BBHs. 
For example, excluding this event from the BBH population leads to inferring a BBH merger rate of $\mathcal{R_{\rm BBH}}=23.9^{+14.9}_{-8.6}\rm Gpc^{-3} yr^{-1}$ while including this event in the analysis would raise the merger rate to 
$\mathcal{R_{\rm BBH}}=58^{+54}_{-29}\rm Gpc^{-3} yr^{-1}$\citep{O3apop}. Similarly, considering this event in the BBH population has large implications on the lower BH mass.

\acknowledgements 
We are thankful to the referee for their constructive comments. 
MTS thanks the Heising-Simons Foundation, the Danish National Research Foundation (DNRF132), and NSF (AST-1911206 and AST-1852393) for support.
DW thanks the NSF (PHY-1912649) for support.
MTS is thankful to Javier Roulet, Ryan Foley, and Enrico Ramirez-Ruiz for useful discussions. We are also thankful to Matthias Kruckow, and Wenbin Lu for comments on our earlier version of the draft.
DW is thankful to Maya Fishbach for valuable feedback.
For their use in our analyses, we would like to acknowledge \textsc{Numpy} and \textsc{Scipy} \citep{scipy}, \textsc{Emcee} \citep{2013PASP..125..306F}, \textsc{Matplotlib} \citep{matplotlib}, \textsc{AstroPy} \citep{astropy:2013,astropy:2018}, and \textsc{h5py} \citep{collete_2013}.

\appendix

\section{Population inference}
\label{app:pop-inf}

To measure the intrinsic, selection bias-free slope shown in Figure \ref{trend-line}, we use a hierarchical Bayesian model, as described in \citet{2019PhRvD.100d3012W}. The population likelihood over the merger rate ($\mathcal{R}$), slope ($A$), and intercept ($B$) is given by
\begin{equation}
  \mathcal{L}(A, B, \mathcal{R}) \propto
  e^{-\mu(A, B, \mathcal{R})} \,
  \prod_{i=1}^{\mathrm{N_{\mathrm{det}}}}
   \int
    \ell_i(m_{1}, m_{2}) \,
    \mathcal{R} \, p(m_1, m_2 | A, B) \,
    \mathrm{d}m_1 \, \mathrm{d}m_2.
\end{equation}

$\mu$ is the average number of detections we would detect during our survey if our candidate population $(A, B, \mathcal{R})$ were correct. We assume a detection is made if it would produce an SNR of 8 in one detector. We approximate an O3a-scale detector by using the Advanced LIGO 140 Mpc range noise curve \citep{T1800545}.

The rest of the likelihood is a product over each of our $N_{\mathrm{det}}$ detections. Each factor in the product is a population-weighted average over the marginalized likelihood for one of our detections, $\ell_i(m_{1}, m_{2})$. Under our chosen population model, described in \S3,
\begin{equation}
  p(m_1, m_2 | A, B) \propto
  \begin{dcases*}
    m_1^{-2.35} \, \delta(m_2 - (A m_1 + B)), & if $m_1 \in [4, 50] \, \msun{}$
    \\
    0, & otherwise.
  \end{dcases*}
\end{equation}
We cannot evaluate the probability density function numerically---since it contains a delta function---but we can still draw samples from it. Therefore, we approximate the integral by Monte Carlo
\begin{equation}
  \int
    \ell_i(m_{1}, m_{2}) \,
    \mathcal{R} \, p(m_1, m_2 | A, B) \,
    \mathrm{d}m_1 \, \mathrm{d}m_2
  \approx
  \frac{1}{N_{\mathrm{samples}}}
  \sum_{j=1}^{N_{\mathrm{samples}}} \mathcal{R} \, \ell_i(m_{1,j}, m_{2,j}),
\end{equation}
where $m_{1,j}$ is drawn from a distribution proportional to $m_1^{-2.35}$, and $m_2 = A m_1 + B$. The LIGO/Virgo collaboration released posterior samples for each detection, but not the marginalized likelihoods that we need. However, the posteriors for GW190814 and GW190426\_152155 are approximately Gaussian, and the priors on $m_1$ and $m_2$ are approximately uniform, so we approximate the marginal likelihood by fitting a Gaussian to the posterior samples, taking the sample mean and covariance.

Now able to evaluate the population likelihood, we use Bayes' theorem to evaluate the posterior, $p(A, B, \mathcal{R} | \mathrm{data}) \propto \pi(A, B, \mathcal{R}) \, \mathcal{L}(A, B, \mathcal{R})$. We assume the prior is uniform over all three parameters. Finally, we draw samples from this posterior distribution, using the Goodman and Weare's affine invariant Markov chain Monte Carlo (MCMC) ensemble sampler \citep{goodman2010ensemble}, with the \textsc{Emcee} Python package \citep{2013PASP..125..306F}.

\bibliographystyle{yahapj}
\bibliography{the_entire_lib}
\end{document}